\documentclass[reqno, 11pt]{amsart}
\usepackage{amsfonts,latexsym}
\usepackage{amsmath}
\usepackage{amscd}
\usepackage{float,amsmath,amssymb,mathrsfs,bm,multirow,graphics}
\usepackage[dvips]{graphicx}
\usepackage[percent]{overpic}

\newcommand{\nequation}{\setcounter{equation}{0}}

\newcommand{\R}{{\Bbb R}}

\newcommand{\C}{{\Bbb C}}

\newcommand{\proofbegin}{\noindent{\it Proof.\,\,}}
\newcommand{\proofend}{\hfill$\Box$\bigskip}

\newtheorem{theorem}{Theorem}[section]
\newtheorem{proposition}[theorem]{Proposition}

\newtheorem{remark}[theorem]{Remark}

\newtheorem{figuretext}{Figure}

\input epsf
\title[Solution of the global relation]{\sc The solution of the global relation for the derivative nonlinear Schr\"odinger equation on the half-line}
\author{Jonatan Lenells}
\address{Department of Mathematics, Baylor University, One Bear Place 97328, Waco, TX 76798, USA.}
\email{Jonatan\_Lenells@baylor.edu}
    
\begin{document}

\begin{abstract} 
\noindent
We consider initial-boundary value problems for the derivative nonlinear Schr\"odinger (DNLS) equation on the half-line $x > 0$. In a previous work, we showed that the solution $q(x,t)$ can be expressed in terms of the solution of a Riemann-Hilbert problem with jump condition specified by the initial and boundary values of $q(x,t)$. However, for a well-posed problem, only part of the boundary values can be prescribed; the remaining boundary data cannot be independently specified, but are determined by the so-called global relation. In general, an effective solution of the problem therefore requires solving the global relation. Here, we present the solution of the global relation in terms of the solution of a system of nonlinear integral equations. This also provides a construction of the Dirichlet-to-Neumann map for the DNLS equation on the half-line. 
\end{abstract}

\maketitle
\tableofcontents

\section{Introduction}\nequation
The derivative nonlinear Schr\"odinger (DNLS) equation
\begin{equation}\label{dnls} 
  iq_t + q_{xx} = i\left(|q|^2q\right)_x,
\end{equation}
where $q(x,t)$ is a complex-valued function, arises in the study of wave propagation in optical fibers \cite{Agrawal2007} and in plasma physics \cite{Mjolhus76} (see \cite{CYL2006} for further references). It is an integrable equation and the initial value problem on the line can be analyzed by means of the Inverse Scattering Transform (IST) as demonstrated by Kaup and Newell \cite{KN1978}. In the last fifteen years, a generalization of the IST to initial-boundary value (IBV) problems developed by Fokas and his collaborators \cite{F1997, F2002, Fbook}, has made it possible to analyze equations such as (\ref{dnls}) on domains involving a boundary. Several of the most well-known integrable PDEs (such as the KdV, modified KdV, nonlinear Schr\"odinger (NLS), sine-Gordon, and Ernst equations) have been investigated using this approach, see e.g. \cite{BFS2006, BS2003, F-I-S, Kamvissis, LFgnls, LFernst, MK2006, P2005}. 

Recently, the Fokas method was implemented to the DNLS equation (\ref{dnls}) posed on the half-line $x > 0$ \cite{Ldnls}.\footnote{Physically, this type of initial-boundary value problem arises naturally. For example, assuming that we can create or measure the waves at some fixed point $x = 0$ in space, we arrive at an initial-boundary value problem on the half-line with initial data given by the initial wave profile for $x >0$ and boundary data provided by the measurements at $x = 0$.}
Provided that the solution exists, it was shown in \cite{Ldnls} that the solution $q(x,t)$ can be recovered from the initial and boundary data via the solution of a $2 \times 2$-matrix Riemann--Hilbert (RH) problem. The jump matrix for this RH problem is given explicitly in terms of four spectral functions $a(k)$, $b(k)$, $A(k)$, and $B(k)$, where $k \in \C$ is the spectral parameter of the associated Lax pair. The functions $a(k)$ and $b(k)$ are defined in terms of the initial data $q_0(x) = q(x,0)$ via a system of linear Volterra integral equations. The functions $A(k)$ and $B(k)$ are defined in terms of the boundary data $g_0(t) = q(0,t)$ and $g_1(t) = q_x(0,t)$ also via a system of linear Volterra integral equations. 
However, for a well-posed problem, only one of the functions $g_0$ and $g_1$ (or a combination of these two functions) can be specified; the remaining boundary data cannot be independently specified, but are determined by the so-called global relation.
Thus, before the functions $A(k)$ and $B(k)$ can be constructed from the above linear integral equations, the global relation must first be used to eliminate the unknown boundary data.

The analysis of the global relation can take place in two different domains: in the physical domain or in the spectral domain. Although these two domains are related by a transform, each viewpoint has its own advantages. 

In the first part of this paper, we analyze the global relation in the spectral domain. We will show that $A(k)$ and $B(k)$ can be determined via the solution of a system of nonlinear integral equations formulated explicitly in terms of the initial data and the known boundary values.

In the second part of the paper, we analyze the global relation in the physical domain. We use a Gelfand-Levitan-Marchenko (GLM) representation to derive an expression for the generalized Dirichlet-to-Neumann map (i.e. the map which determines the unknown boundary values from the known ones). Once the unknown boundary values have been determined, $A(k)$ and $B(k)$ can be constructed from the linear integral equations mentioned earlier. This representation has the numerical advantage that the system of integral equations is defined on a bounded domain. 

In the case of the NLS equation, a construction of the Dirichlet-to-Neumann map was presented in \cite{BFS2003, F2005}. The Dirichlet-to-Neumann map for the sine-Gordon equation as well as the two versions of mKdV were analyzed in \cite{F2005}. The analysis of the KdV equation presents some novel difficulties which were finally overcome in \cite{TF2008}. 
Although our approach is inspired by the preceeding references, the analysis here presents a number of novelties: (a) The solution of the global relation presented in section \ref{spectralsolutionsec} takes place entirely in the spectral domain. This is in contrast to the approach of \cite{F2005}, in which it is necessary to derive a GLM representation for an appropriate eigenfunction of the Lax pair before an analogous result can be obtained. The introduction of a GLM representation amounts to passing from the spectral to the physical plane. Here we remain in the spectral plane throughout the derivation, which simplifies the arguments both on a practical and on a conceptual level. The observation that it is possible to solve the global relation directly in the spectral plane was first made for the NLS equation in \cite{FLpreprint}. (b) In the special, but important, case of vanishing initial data $q_0 = 0$, we will show that a further analysis of the global relation can be used to simplify the integral equations derived in the physical plane considerably. In fact, the number of equations reduces from five to two in this case. The case of vanishing initial data is of special interest in applications, where it can sometimes be assumed that the physical field under consideration is initially at rest before exteriorly created waves enter the domain. This is also the relevant initial condition in the application of the numerical methods based on the so-called ``exact nonreflecting boundary conditions'' \cite{Z2006}. A similar simplification for the NLS equation was presented in \cite{FLpreprint}. (c) The Lax pair for equation (\ref{dnls}) contains terms of higher order in the spectral parameter $k$ than what is the case for the examples above. However, by utilizing additional symmetries, calculations can still be kept at a reasonable length.
(d) As noted in \cite{Ldnls}, the definition of eigenfunctions with the appropriate asymptotics of the Lax pair of (\ref{dnls}) naturally leads to the introduction of a certain closed two-form $\Delta$. In order to close the system of integral equations which characterizes the solution of the global relation, the system must be supplemented by an additional (somewhat complicated) equation for the second component of $\Delta$.

The paper is organized as follows: In section \ref{laxsec}, we recall the Lax pair formulation and the global relation associated with (\ref{dnls}). In sections \ref{spectralsolutionsec} and \ref{physicalsolutionsec}, the global relation is analyzed in the spectral and physical domains, respectively. In section \ref{GLMsec}, we derive a GLM representation for an appropriate eigenfunction of the Lax pair.

\section{A Lax pair and the global relation}\nequation\label{laxsec}
Equation (\ref{dnls}) admits the Lax pair \cite{KN1978}
\begin{equation}\label{Psilax}
\begin{cases}
	& \Psi_x + ik^2[\sigma_3, \Psi] = U_1 \Psi, \\
	& \Psi_t + 2ik^4[\sigma_3, \Psi] = U_2 \Psi,
\end{cases}
\end{equation}
where $\sigma_3 = \text{diag}(1,-1)$, $k \in \hat{\C} = \C \cup \{\infty\}$ is a spectral parameter, $\Psi(x,t,k)$ is a $2\times2$-matrix valued eigenfunction, and the $2\times2$-matrix valued functions $Q(x,t)$ and $\{U_j(x,t,k)\}_1^2$ are defined by
\begin{align*}
&Q = \begin{pmatrix} 0 & q \\ \bar{q} & 0 \end{pmatrix}, \qquad 
U_1 = k Q, \qquad U_2 = 2k^3 Q -ik^2Q^2\sigma_3 - ik Q_x\sigma_3 + k Q^3.
\end{align*}
Following \cite{Ldnls}, we define a transformed eigenfunction $\mu(x,t,k)$ by 
\begin{equation}\label{Psimurelation}
\Psi(x,t,k) = e^{i\int^{(x,t)}_{(0, 0)} \Delta \sigma_3}\mu(x,t,k) e^{i\int^{(0,0)}_{(\infty, 0)} \Delta \sigma_3},
\end{equation}
where $\Delta = \Delta_1dx + \Delta_2 dt$ is the closed real-valued one-form
\begin{align}\label{Deltadef}  
  \Delta(x,t)& = \frac{1}{2}|q|^2dx + \left(\frac{3}{4} |q|^4 - \frac{i}{2}(\bar{q}_xq - \bar{q}q_x)\right)dt.
  \end{align}
We introduce $\mathcal{Q}(x,t)$ and $\mathcal{Q}_1(x,t)$ by 
$$\mathcal{Q} = e^{-i\int^{(x,t)}_{(0, 0)} \Delta \hat{\sigma}_3}Q, \qquad 
\mathcal{Q}_1 = e^{-i\int^{(x,t)}_{(0, 0)} \Delta \hat{\sigma}_3}Q_x,$$
where $\hat{\sigma}_3$ acts on a $2\times 2$ matrix $A$ by $\hat{\sigma}_3A = [\sigma_3, A]$, i.e. $e^{\hat{\sigma}_3}A = e^{\sigma_3} A e^{-\sigma_3}$.
The function $\mu$ satisfies the Lax pair
\begin{equation}\label{mulax}  
\begin{cases}
	& \mu_x + ik^2 [\sigma_3, \mu] = V_1\mu, \\
	& \mu_t + 2ik^4 [\sigma_3, \mu] = V_2\mu,
\end{cases}
\end{equation}
where the $2 \times 2$-matrix valued functions $\{V_j(x,t,k)\}_1^2$ are given by
\begin{equation}\label{V1V2def}
V_1 = k\mathcal{Q} - i \Delta_1 \sigma_3, \qquad V_2 
= 2k^3 \mathcal{Q} - ik^2\mathcal{Q}^2\sigma_3 - ik \mathcal{Q}_1 \sigma_3 + k \mathcal{Q}^3 -i\Delta_2 \sigma_3.
\end{equation}

Let $T > 0$ be some given final time; we will assume that $T < \infty$. Assume that $q(x,t)$ is a solution of (\ref{dnls}) with sufficient smoothness in the domain $\{0 < x < \infty, 0 < t < T\}$ and with sufficient decay as $x \to \infty$. We introduce three solutions $\{\mu_j(x,t,k)\}_1^3$ of (\ref{mulax}) as the solutions of the following linear Volterra integral equations:
\begin{equation}\label{mujdef}  
  \mu_j(x,t,k) = I + \int_{(x_j, t_j)}^{(x,t)} e^{i[k^2(x'-x) + 2k^4(t' - t)]\hat{\sigma}_3}\left((V_1\mu_j)(x',t',k) dx' + (V_2 \mu_j) (x',t',k) dt'\right),
\end{equation}
where $(x_1, t_1) = (0, T)$, $(x_2, t_2) = (0, 0)$, $(x_3, t_3) = (\infty, t)$, and $I$ denotes the $2\times 2$ identity matrix. 
The eigenfunctions $\{\mu_j\}_1^3$ satisfy the symmetries 
\begin{align}\label{mujsymm}
  \mu_j(x,t,k) = \sigma_1\overline{\mu_j(x,t,\bar{k})}\sigma_1,  \qquad \mu_j(x,t,k) = \sigma_3\mu_j(x,t,-k)\sigma_3, \qquad j= 1,2,3.
\end{align}
The first of these symmetries implies that we can introduce complex-valued functions $a(k)$, $b(k)$, $A(k)$, $B(k)$, $\Phi_1(t, k)$, $\Phi_2(t, k)$ by
\begin{align}\label{abABPhi12def}
& s(k) = \begin{pmatrix} \overline{a(\bar{k})} & b(k) \\ \overline{b(\bar{k})} & a(k) \end{pmatrix}, \qquad
S(k) = \begin{pmatrix} \overline{A(\bar{k})} & B(k) \\ \overline{B(\bar{k})} & A(k) \end{pmatrix}, 
	\\ \nonumber
& \mu_2(0,t,k) =  \begin{pmatrix} \overline{\Phi_2(t, \bar{k})} & \Phi_1(t, k)e^{-2i\int_{(0,0)}^{(0,t)} \Delta } \\ \overline{\Phi_1(t, \bar{k})}e^{2i\int_{(0,0)}^{(0,t)} \Delta} & \Phi_2(t, k) \end{pmatrix},
\end{align}
where $s(k)$ and $S(k)$ are defined by
\begin{equation}\label{sSdef}  
  s(k) = \mu_3(0,0,k), \qquad S(k) = [e^{2ik^4T\hat{\sigma}_3}\mu_2(0,T,k)]^{-1}.
\end{equation}
Since $a(k) = a(-k)$, the zeros of $a(k)$ always come in pairs; if $k_j$ is a zero, then so is $-k_j$. 

The function $\mu_3(x,0,k)$, and hence also $s(k)$, can be constructed from the initial data $q_0(x)$ via the linear Volterra integral equation
\begin{equation}\label{mu3x0def}  
  \mu_3(x,0,k) = I + \int_\infty^x e^{ik^2(x' - x)\hat{\sigma}_3} (V_1\mu_3)(x', 0, k) dx'.
\end{equation}  
Similarly, $\mu_2(0,t,k)$, and hence also $S(k)$, can be constructed from the boundary values $g_0(t)$ and $g_1(t)$ via the linear Volterra integral equation
\begin{equation}\label{mu20tdef}  
  \mu_2(0, t, k) = I + \int_0^t e^{2ik^4(t' - t)\hat{\sigma}_3} (V_2\mu_2)(0, t', k) dt'.
\end{equation}

\begin{figure}
\begin{center}
    \includegraphics[width=.5\textwidth]{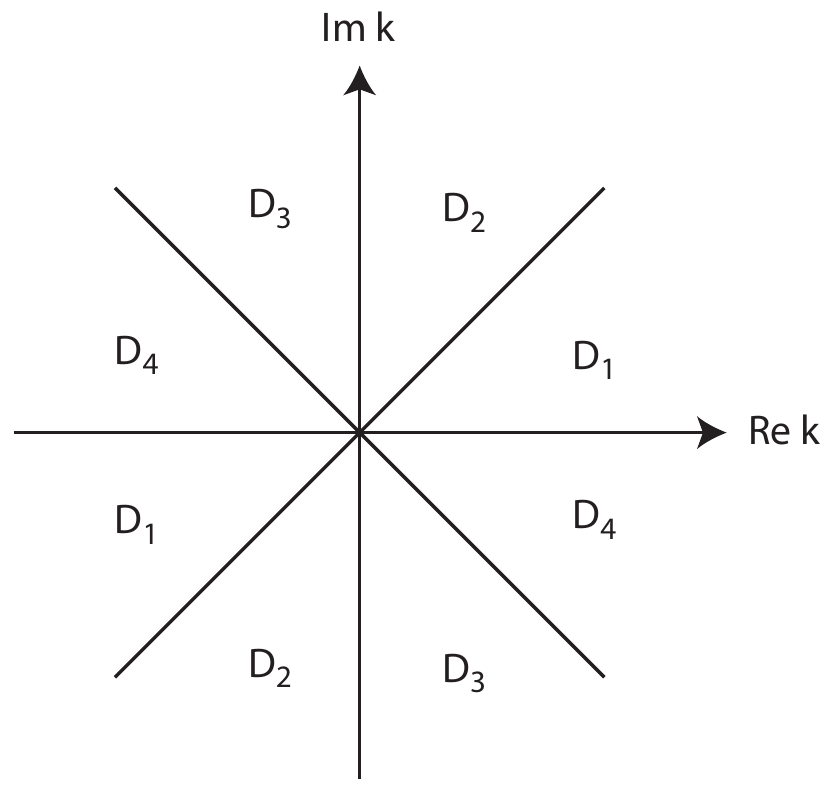} \\
     \begin{figuretext}\label{Djs.pdf}
        The sets $\{D_j\}_1^4$ that decompose the complex $k$-plane.
     \end{figuretext}
     \end{center}
\end{figure}

Defining the sets $D_j$, $j = 1, \dots, 4$, by (see figure \ref{Djs.pdf})
$$D_j = \left\{(j-1)\pi/2 < \arg(k^2) < j\pi/2\right\},$$
it follows from (\ref{mu3x0def})-(\ref{mu20tdef}) that $a(k)$ and $b(k)$ are analytic and bounded in $D_1 \cup D_2$, while $A(k)$ and $B(k)$ are entire functions which are bounded in $D_1 \cup D_3$. The functions $\Phi_1(t, k)$ and $\Phi_2(t, k)$ are entire functions of $k$ which are bounded forÊ $k \in D_2 \cup D_4$.

\subsection{The global relation}
The eigenfunctions $\mu_2$ and $\mu_3$ are related by
\begin{equation}\label{preGR}
  \mu_3(x,t,k) = \mu_2(x,t,k)e^{-i(k^2x + 2k^4t)\hat{\sigma}_3}s(k),
\end{equation}  
where the first and second columns are valid for $k \in \bar{D}_3 \cup \bar{D}_4$ and $k \in \bar{D}_1 \cup \bar{D}_2$, respectively.
The relation obtained by evaluating (\ref{preGR}) at $(0,t)$ is called the {\it global relation}. 
It imposes a relation between the Dirichlet and Neumann boundary values of $q(x,t)$. The Dirichlet-to-Neumann map is determined by solving this relation for the unknown boundary values. Defining functions $c(t, k)$ and $d(t, k)$ by 
$$c(t, k) = \frac{e^{2i\int_{(0,0)}^{(0,t)} \Delta}}{a(k)}(\mu_3(0,t,k))_{12}, \qquad d(t, k) = \frac{1}{a(k)}(\mu_3(0,t,k))_{22},$$
we can write the $(12)$ and $(22)$ entries of the global relation as 
\begin{subequations}\label{GRcd}
\begin{align}\label{GRc}
& c(t, k) = \Phi_1(t, k) + e^{2i\int_{(0,0)}^{(0,t)} \Delta} \frac{b(k)}{a(k)}\overline{\Phi_2(t, \bar{k})} e^{-4ik^4t}, \qquad k \in \bar{D}_1 \cup \bar{D}_2, 
	\\ \label{GRd}
& d(t, k) = \Phi_2(t, k) + e^{2i\int_{(0,0)}^{(0,t)} \Delta} \frac{b(k)}{a(k)}\overline{\Phi_1(t, \bar{k})} e^{-4ik^4t}, \qquad k \in \bar{D}_1 \cup \bar{D}_2.
\end{align}
\end{subequations}
We find from the definition of $\mu_3$ that the functions $c(t, k)$ and $d(t, k)$ are analytic and bounded for $k \in D_1 \cup D_2$ away from the possible zeros of $a(k)$.

\section{The solution of the global relation: spectral domain}\nequation\label{spectralsolutionsec}
In this section we analyze the global relation in the spectral plane. For concreteness, we consider the Dirichlet problem; the treatment of the Neumann problem is analogous. The main result is theorem \ref{th1}, which expresses the spectral functions $A(k)$ and $B(k)$ in terms of the Dirichlet boundary data and the initial data via the solution of a system of nonlinear integral equations. 

\begin{figure}
\begin{center}
 \begin{overpic}[width=.6\textwidth]{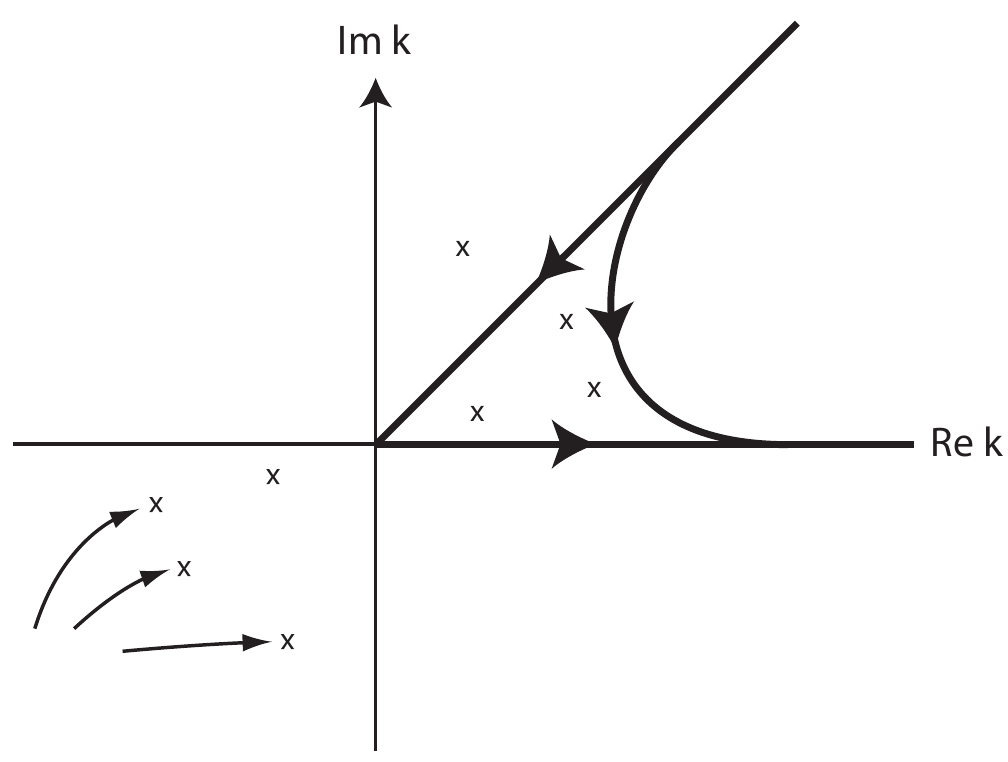}
      \put(52,54){\large $\gamma$}
      \put(64,43){\large $\gamma'$}
      \put(77, 49){\large $D_1^+$}
      \put(-15.8,8.2){\large zeros of $a(k)$}
    \end{overpic}
     \begin{figuretext}\label{gamma.pdf}
        The contour $\gamma$ in the complex $k$-plane and the deformed contour $\gamma'$. The zeros of $a(k)$ are symmetrically distributed with respect to the origin.
     \end{figuretext}
     \end{center}
\end{figure}

For a subset $\Sigma \subset \C$ of the complex $k$-plane, we let $\partial \Sigma$ denote the boundary of $\Sigma$, oriented so that $\Sigma $ lies to the left of $\partial \Sigma $. We let $D_j^+ = D_j \cap \{\text{Im}\,k \geq 0\}$, $j = 1, \dots, 4$. Moreover, we let $\gamma$ denote the contour $\partial D_1^+$ and let $\gamma'$ denote a contour obtained from Ê$\gamma$ by deforming it so that it does not surround the possible zeros of $a(k)$, see figure  \ref{gamma.pdf}.

\begin{theorem}\label{th1}
Let $T < \infty$. Let $q_0(x)$, $x \geq 0$, be a function of Schwartz class and let $g_0(t)$, $0 \leq t < T$, be a sufficiently smooth function. Assume that $q_0$ and $g_0$ are compatible at $x = t = 0$ in the sense that $q_0(0) = g_0(0)$. Let $a(k)$ and $b(k)$ be defined for $k \in \bar{D}_1 \cup \bar{D}_2$ by equations (\ref{abABPhi12def})-(\ref{mu3x0def}). Suppose that $a(k)$ has a finite (possibly empty) set of simple zeros. We denote these zeros by $\{k_j, -k_j\}_1^{N}$ and assume that $\{k_j\}_1^{N} \subset D_1$. 

Then the spectral functions $A(k)$ and $B(k)$ defined by (\ref{abABPhi12def})-(\ref{sSdef}) are given by
\begin{align}\label{ABexpressions}
& A(k)  = \overline{\Phi_2(T, \bar{k})} 
\qquad B(k) = -\Phi_1(T, k)e^{4ik^4T}e^{-2i\int_{0}^T \Delta_2(0,t)dt},
\end{align}
where the complex-valued functions $\Phi_1(t, k)$, $\Phi_2(t, k)$ and the real-valued function $\Delta_2(0,t)$ satisfy the following system of nonlinear integral equations:
\begin{subequations}\label{PhiDeltaeqs}
\begin{align}\label{Phieqs}
&\begin{pmatrix} \Phi_1(t,k) \\ \Phi_2(t, k) \end{pmatrix}
= \begin{pmatrix} 0 \\ 1 \end{pmatrix}
	\\ \nonumber
&+ \int_0^t \begin{pmatrix} e^{4ik^4(t' - t)}\left[(-i|g_0|^2k^2 + i\Delta_2) \Phi_{1} + k \left(|g_0|^2 g_0+2 k^2 g_0+i g_1\right) \Phi_{2}\right](t',k)\\
\left[k \left(|g_0|^2 \bar{g}_0+2 k^2 \bar{g}_0- i \bar{g}_1\right)\Phi_{1} + (i|g_0|^2k^2 + i\Delta_2) \Phi_{2}\right](t',k) 
\end{pmatrix}  dt',
	\\  \label{Delta2eq}  
&  \Delta_2(0,t) =  \frac{1}{4}|g_0(t)|^4 - 4 \text{\upshape Im}(\bar{g}_0(t)c^{(3)}(t)) 
	\\ \nonumber
& \qquad \qquad \ - 2 |g_0(t)|^2 \text{\upshape Im}\left[\frac{2}{\pi} \int_{\gamma} k \left[\Phi_2(t, k) - \Phi_2(t, ik)\right]dk\right],
\end{align}
\end{subequations}
where the functions $g_1(t)$ and $c^{(3)}(t)$ are given for $0 \leq t < T$ by
\begin{align} \label{g1expression}
& g_1(t) = \frac{i}{2} |g_0(t)|^2g_0(t)  + 4 c^{(3)}(t)
+ \frac{4g_0(t)}{\pi} \int_{\gamma} k \left[\Phi_2(t, k) - \Phi_2(t, ik)\right]dk,
	\\  \label{c3expression}
&c^{(3)}(t) =  \frac{i}{\pi}\int_{\gamma}k^2\left(\Phi_1(t,k) + i\Phi_1(t, ik) + \frac{i g_0(t)}{k}\right) dk
	\\ \nonumber
& \;+ e^{2i\int_0^t \Delta_2(0,t')dt'}\left[ \frac{2i}{\pi }\int_{\gamma} k^2 \frac{b(k)}{a(k)} \overline{\Phi_2(t, \bar{k})} e^{-4ik^4 t}dk
 + 4 \sum_{k_j \in D_1^+} k_j^2 \frac{b(k_j)}{\dot{a}(k_j)} \overline{\Phi_2(t, \bar{k}_j)}e^{-4ik_j^2 t}\right].
 \end{align}
\end{theorem}
\proofbegin
Letting $g_1(t) := q_x(0,t)$, equations (\ref{ABexpressions}) and (\ref{Phieqs}) follow from the definition of $\Phi_1$ and $\Phi_2$ together with (\ref{mulax}).
In order to derive (\ref{g1expression}), we substitute the expansion 
$$\mu_j = I + \frac{\mu_j^{(1)}}{k} + \frac{\mu_j^{(2)}}{k^2} + \frac{\mu_j^{(3)}}{k^3} + O\biggl(\frac{1}{k^4}\biggr), \qquad k \to \infty, \qquad j = 1,2,3,$$ 
into (\ref{mulax}) and solve for the $\mu_j^{(n)}$'s. By the second symmetry in (\ref{mujsymm}), $\mu_j^{(n)}$ is a diagonal (off-diagonal) matrix for $n$ even (odd). We find (see \cite{Ldnls} for further details of this type of argument)
$$\mu_j^{(1)} = -\frac{i}{2}\sigma_3\mathcal{Q}, \qquad 2i\sigma_3\mu_j^{(3)} = \mathcal{Q}\mu_j^{(2)} + \frac{1}{4}\mathcal{Q}^3 + \frac{i}{2}\sigma_3\mathcal{Q}_1,$$
and
$$\mu_{jx}^{(2)} = \frac{1}{4} QQ_x + \frac{i}{8}Q^4\sigma_3,
\qquad \mu_{jt}^{(2)} = \frac{i}{4}Q^6\sigma_3 + \frac{3}{4}Q^3Q_x + \frac{i}{4}Q_x^2\sigma_3 - \frac{i}{4}QQ_{xx}\sigma_3 .$$
Thus, as $k \to \infty$,
\begin{align}\nonumber
 & e^{2i\int_{(0,0)}^{(x,t)} \Delta}(\mu_j(x,t,k))_{12} = -\frac{iq}{2k}  + \frac{1}{k^3}\left(\frac{q_x}{4} -\frac{iq}{2}\int_{(x_j, t_j)}^{(x,t)} \omega
 - \frac{i}{8}|q|^2 q\right) + O \biggl(\frac{1}{k^5} \biggr), 
  	\\ \label{mujasymptotics}
 &  (\mu_j(x,t,k))_{22} = 1 +  \frac{1}{k^2}\int_{(x_j, t_j)}^{(x,t)} \omega + O \biggl(\frac{1}{k^4} \biggr), 
\end{align}
where the closed one-form $\omega$ is defined by
$$\omega = \left(\frac{1}{4} \bar{q}q_x - \frac{i}{8} |q|^4\right)dx - \frac{i}{4}\left(|q|^6 + 3i |q|^2\bar{q} q_x + |q_x|^2 - \bar{q}q_{xx} \right) dt.$$
The expansions in (\ref{mujasymptotics}) are valid for $k$ approaching $\infty$ within the regions of boundedness of $(\mu_j)_{12}$ and $(\mu_j)_{22}$. 
Hence,
\begin{align}\nonumber
& c(t, k) = - \frac{ig_0(t)}{2k} + \frac{c^{(3)}(t)}{k^3} + O \biggl(\frac{1}{k^5} \biggr), \qquad k \to \infty, \quad k \in D_1 \cup D_2,
	\\ \label{cPhiexpansions}
& \Phi_1(t, k) = - \frac{ig_0(t)}{2k} + \frac{c^{(3)}(t)}{k^3} + O \biggl(\frac{1}{k^5} \biggr), \qquad k \to \infty, \quad k \in D_2 \cup D_4,
	\\ \nonumber
& \Phi_2(t, k) = 1 +  \frac{\Phi_2^{(2)}(t) }{k^2}+ O \biggl(\frac{1}{k^4} \biggr), \qquad k \to \infty, \quad k \in D_2 \cup D_4,
\end{align}
where
$$c^{(3)}(t) = \frac{g_1}{4} -\frac{ig_0}{2}\int_{(0,0)}^{(0,t)} \omega - \frac{i}{8}|g_0|^2g_0, \qquad
\Phi_2^{(2)}(t) = \int_{(0,0)}^{(0,t)} \omega.$$
It follows that
\begin{equation}\label{g1cPhi}
g_1(t) = \frac{i}{2} |g_0|^2g_0 + 4c^{(3)} + 2ig_0 \Phi_2^{(2)}, \qquad 0 \leq t < T.
\end{equation}

The function $\Phi_2(t, k) - \Phi_2(t, ik)$ is an entire function of $k$ and satisfies
$$\Phi_2(t, k) - \Phi_2(t, ik) = \frac{2}{k^2} \Phi_2^{(2)}(t) + O\biggl(\frac{1}{k^4}\biggr), \qquad k \to \infty, \quad k \in D_2 \cup D_4.$$
Letting $C_R$, $R > 0$, denote the large arc of radius $R$ which closes $D_2^+$ at infinity, 
$$C_R = \{Re^{i\alpha}Ê\in \C \; |\; \pi/4 \leq \alpha \leq \pi/2\},$$
we find
$$\int_{\partial D_2^+} k \left[\Phi_2(t, k) - \Phi_2(t, ik)\right] dk
= - \lim_{R \to \infty} \int_{C_R} k \left[\Phi_2(t, k) - \Phi_2(t, ik)\right] dk
= -\frac{i\pi}{2}\Phi_2^{(2)}(t).$$
Consequently,
\begin{equation}\label{Phi22expression}  
  \Phi_2^{(2)}(t) = -\frac{2}{\pi i } \int_{\partial D_2^+} k \left[\Phi_2(t, k) - \Phi_2(t, ik)\right]dk.
\end{equation}
A change of variables $k \to ik$ in the part of the integration that runs along the imaginary axis shows that the contour $\partial D_2^+$ in (\ref{Phi22expression}) can be replaced by $-\gamma$. Thus, equations (\ref{g1cPhi}) and (\ref{Phi22expression}) yield (\ref{g1expression}).

Substituting the expression (\ref{g1cPhi}) for $g_1$ into the definition (\ref{Deltadef}) of $\Delta$ evaluated at $(0,t)$, we find
$$\Delta_2(0,t) = \frac{1}{4}|g_0|^4 - 4 \text{Im}\bigl(\bar{g}_0c^{(3)}\bigr) - 2 |g_0|^2 \text{Re}\,\Phi_2^{(2)}.$$
In view of (\ref{Phi22expression}), this yields (\ref{Delta2eq}).

It remains to derive the expression (\ref{c3expression}) for $c^{(3)}(t)$. Assume first that $a(k)$ has no zeros.
Let $\epsilon > 0$. The function 
$$k\left[c(t, k) + \frac{ig_0(t)}{4} \left(\frac{1}{k - \epsilon e^{- \pi i/4}} + \frac{1}{k + \epsilon e^{- \pi i/4}}\right)\right]$$
is analytic in $D_1 \cup D_2$ and of $O(1/k^2)$ as $k \to \infty$. Thus, the sectionally analytic function
$$\begin{cases}  k\left[c(t, k) + \frac{ig_0(t)}{4} \left(\frac{1}{k - \epsilon e^{- \pi i/4}} + \frac{1}{k + \epsilon e^{- \pi i/4}}\right)\right], \qquad k \in \bar{D}_1 \cup \bar{D}_2, \\
-ik\left[c(t, ik) + \frac{ig_0(t)}{4} \left(\frac{1}{ik - \epsilon e^{- \pi i/4}} + \frac{1}{ik + \epsilon e^{- \pi i/4}}\right)\right], \qquad k \in \bar{D}_3 \cup \bar{D}_4, 
 \end{cases}$$
satisfies a scalar RH problem with jump across $\R \cup i\R$. Let $\Gamma$ denote the oriented contour
$$\Gamma = [-i\infty, 0] \cup [0, -\infty] \cup [i\infty, 0] \cup [0, \infty].$$ 
We find
\begin{align}\label{cRHsolution}
&k\left[c(t, k) + \frac{ig_0(t)}{4} \left(\frac{1}{k - \epsilon e^{- \pi i/4}} + \frac{1}{k + \epsilon e^{- \pi i/4}}\right)\right]
	\\ \nonumber
& = \frac{1}{2\pi i} \int_\Gamma \left[lc(t, l) + ilc(t, il) 
+ \frac{ig_0(t)}{4}\sum_{s=0}^3 \frac{i^sl}{i^sl - \epsilon e^{-\pi i/4}}\right]\frac{dl}{l - k}, \qquad k \in D_1 \cup D_2.
\end{align}
Thus,
$$c^{(3)}(t) + \frac{g_0(t)}{2}\epsilon^2= - \frac{1}{2\pi i}\int_\Gamma k\left[kc(t, k) + ikc(t, ik) + \frac{ig_0(t)}{4}\sum_{s=0}^3 \frac{i^sk}{i^sk - \epsilon e^{-\pi i/4}}\right]dk.$$
Letting $\epsilon \to 0$ and using the global relation (\ref{GRc}) to eliminate $c(t, k)$ and $c(t, ik)$ from the integrand, we find
\begin{align}\label{c3midstep}
c^{(3)}(t) = &-\frac{1}{2\pi i}\int_\Gamma k^2\left[\Phi_1(t, k) + i\Phi_1(t, ik) + \frac{i g_0(t)}{k}\right]dk
	\\ \nonumber
& -  \frac{e^{2i\int_{(0,0)}^{(0,t)} \Delta}}{2\pi i}\int_\Gamma k^2 \left[\frac{b(k)}{a(k)} \overline{\Phi_2(t, \bar{k})}  + i \frac{b(ik)}{a(ik)} 
\overline{\Phi_2(t, \overline{ik})}\right]e^{-4ik^4 t} dk.
\end{align}
The integrand in the first term on the right-hand side is analytic in $D_2$ and of $O(1/k^3)$ as $kÊ\to \infty$ in $D_2$. Thus, the contour of integration $\Gamma = \partial D_1 + \partial D_2$ in this term can be replaced by $\partial D_1$. 
Using the change of variables $k \to -ik$, we find that the second term on the right-hand side of (\ref{c3midstep}) equals
\begin{equation}\label{secondintsimple}
  -\frac{e^{2i\int_{(0,0)}^{(0,t)} \Delta}}{\pi i}\int_\Gamma k^2 \frac{b(k)}{a(k)} \overline{\Phi_2(t, \bar{k})} e^{-4ik^4 t}dk.
\end{equation}
The global relation (\ref{GRc}) together with the asymptotics (\ref{cPhiexpansions}) imply that the integrand in (\ref{secondintsimple}) is of $O(1/k^3)$ as $k \to \infty$, $k \in D_2$. Thus, the contour of integration in (\ref{secondintsimple}) can be replaced by $\partial D_1$. We arrive at
\begin{align}\label{ccoeff2}
c^{(3)}(t) =&\; -\frac{1}{2\pi i}\int_{\partial D_1}k^2\left[\Phi_1(t,k) + i\Phi_1(t, ik) + \frac{i g_0(t)}{k}\right] dk
	\\ \nonumber
& -\frac{e^{2i\int_{(0,0)}^{(0,t)} \Delta}}{\pi i}\int_{\partial D_1} k^2 \frac{b(k)}{a(k)} \overline{\Phi_2(t,\bar{k})} e^{-4ik^4 t}dk.
\end{align}
The symmetry properties (\ref{mujsymm}) imply that both integrands in (\ref{ccoeff2}) are odd functions of $k$. Thus, we obtain
\begin{align*}\label{}
c^{(3)}(t) = &\; \frac{i}{\pi}\int_{\gamma}k^2\left[\Phi_1(t,k) + i\Phi_1(t, ik) + \frac{i g_0(t)}{k}\right] dk
	\\
& + \frac{2ie^{2i\int_{(0,0)}^{(0,t)} \Delta}}{\pi}\int_{\gamma} k^2 \frac{b(k)}{a(k)} \overline{\Phi_2(t, \bar{k})} e^{-4ik^4 t}dk.
\end{align*}
This proves (\ref{c3expression}) in the case when $a(k) \neq 0$. 

If $a(k)$ has a finite number of simple zeros $\{k_j, -k_j\}_1^{N} \subset D_1 \cup D_2$, we replace equation (\ref{cRHsolution}) with
\begin{align} \nonumber
&k\left[c(t, k) + \frac{ig_0(t)}{4} \left(\frac{1}{k - \epsilon e^{- \pi i/4}} + \frac{1}{k + \epsilon e^{- \pi i/4}}\right)
- \sum_{j = 1}^N c_j(t)\left(\frac{1}{k - k_j} + \frac{1}{k + k_j}\right)\right]
	\\ \nonumber
& = \frac{1}{2\pi i} \int_\Gamma \left[lc(t, l) + ilc(t, il) 
+ \frac{ig_0(t)}{4}\sum_{s=0}^3 \frac{i^sl}{i^sl - \epsilon e^{-\pi i/4}} - \sum_{j=1}^N c_j(t) \sum_{s=0}^3 \frac{i^sl}{i^sl - k_j}\right]\frac{dl}{l - k},
	\\ \label{cRHsolutionreplaced}
& \hspace{10cm} k \in D_1 \cup D_2.
\end{align}
where $c_j(t)$, $j = 1, \dots, N$, denotes the residue of $c(t,k)$ at $k = k_j$, i.e. 
$$c_j(t) = e^{2i\int_{(0,0)}^{(0,t)} \Delta}\frac{b(k_j)}{\dot{a}(k_j)} \overline{\Phi_2(t, \bar{k}_j)}e^{-4ik_j^4 t}.$$
By closing the contour at infinity, we find
$$-\frac{1}{2\pi i} \int_{\Gamma} \sum_{j=1}^N c_j(t)\left[ \sum_{s=0}^3 \frac{i^sl}{i^sl - k_j}\right]\frac{dl}{l - k}
= 2\sum_{j=1}^N c_j + \frac{2}{k^2}\sum_{j = 1}^{N} c_jk_j^2 + O\biggl(\frac{1}{k^4}\biggr)$$
as $k \to \infty$ within $D_1 \cup D_2$. The contribution to $c^{(3)}$ from the term on the right-hand side of (\ref{cRHsolutionreplaced}) that involves the residues $\{c_j\}_1^N$ is therefore $2 \sum_{j = 1}^N c_j k_j^2$. The effect of the term involving the $c_j$'s on the left-hand side of (\ref{cRHsolutionreplaced}) is to add another $2 \sum_{j = 1}^N c_j k_j^2$ to $c^{(3)}(t)$.
Finally, when replacing the contour in (\ref{secondintsimple}) with $\partial D_1$, we now pick up a sum of residues from the zeros of $a(k)$ in $D_2$; this contributes an additional $- 4 \sum_{k_j \in D_2^+} c_j k_j^2$ to $c^{(3)}$. The total net contribution to $c^{(3)}(t)$ from the residues at the zeros of $a(k)$ is therefore
$$2 \sum_{j = 1}^N c_j k_j^2 + 2 \sum_{j = 1}^N c_j k_j^2 - 4 \sum_{k_j \in D_2^+} c_j k_j^2 = 4 \sum_{k_j \in D_1^+} c_j k_j^2,$$
which proves (\ref{c3expression}) also when $a(k) \neq 0$.
\proofend

In \cite{Ldnls}, the solutionÊ $q(x,t)$ was presented in terms of the solution of a matrix RH problem with jump condition specified by the spectral functions $a(k), b(k)$, $A(k)$, and $B(k)$.
Substitution of the expressions (\ref{g1expression}) and (\ref{c3expression}) into (\ref{PhiDeltaeqs}) yields a system of nonlinear integral equations involving the functions $\Phi_1(t, k)$, $\Phi_2(t,k)$, and $\Delta_2(0,t)$. Assuming that this system has a unique solution, $A(k)$ and $B(k)$ can be determined from (\ref{ABexpressions}).

When $a(k) \neq 0$, the functions $A(k)$ and $B(k)$ only enter the formulation of the RH problem as the combination $A(k)/B(k)$ for $k \in \partial D_3$. Then it is sufficient to consider the system satisfied by $\Phi_1(t, k)$, $\Phi_2(t,k)$, and $\Delta_2(0,t)$ for $0 \leq t < T$ and $k \in \partial D_1 \cup \partial D_3$; the function $A(k)/B(k)$, $k \in \partial D_3$, can be recovered from the solution of this system. In the case when $a(k)$ has zeros, the residue conditions also need to be taken into account.

\begin{remark}\upshape\label{intremark}
1. By replacing in the right-hand side of (\ref{c3expression}) the contour $\gamma$ by the contour $\gamma'$ which does not surround the zeros of $a(k)$ (see figure \ref{gamma.pdf}), the sum over $k_j \in D_1^+$ can be absorbed into the preceding integral.

2. The integrals along $\gamma$ which appear on the right-hand sides of (\ref{Delta2eq})-(\ref{c3expression}) are, in general, not absolutely convergent; they are defined as the limits as $R \to \infty$ of the integrals obtained by replacing $\gamma$ with $\gamma_R$, where $\gamma_R = [Re^{\pi i/4}, 0] \cup [0, R]$.
\end{remark}

\section{The GLM representation}\label{GLMsec} \nequation
In this section we derive Gelfand-Levitan-Marchenko (GLM) representations for the eigenfunctions $\{\Phi_j\}_1^2$. These representations will be needed in section \ref{physicalsolutionsec}. 

\begin{proposition}
Suppose $q(x,t)$ is a solution of (\ref{dnls}) with sufficiently smooth boundary data $g_0(t) = q(0,t)$ and $g_1(t) = q_x(0,t)$, $0 \leq t < T$.
The functions $\Phi_1$ and $\Phi_2$ defined in (\ref{abABPhi12def}) can be represented as
\begin{align}\label{GLMrep}
  \begin{pmatrix} \Phi_{1}(t,k) \\ \Phi_2(t,k) \end{pmatrix}
   = \begin{pmatrix} 0 \\ 1 \end{pmatrix} 
   + \int_{-t}^t \begin{pmatrix} I_1(t,s)k + I_3(t,s) k^3 \\ 
    I_0(t,s) + I_2(t,s)k^2\end{pmatrix} e^{2ik^4(s -t)} ds,
\end{align}
where
\begin{subequations}\label{I0123def}
\begin{align}
& I_0(t,s) = j(t,s) + \frac{i}{2}\bar{g}_0(t)l(t,s) + \left(\frac{i}{8}|g_0(t)|^2\bar{g}_0(t) + \frac{1}{4}\bar{g}_1(t)\right)n(t,s), 
	\\\label{I1def}
& I_1(t,s) =  l(t,s) - \frac{i}{2}g_0(t)m(t,s),
	\\
& I_2(t,s) = m(t,s) + \frac{i}{2}\bar{g}_0(t)n (t,s),
	\\
& I_3(t,s) = n(t,s),
\end{align}
\end{subequations}
and the functions $\{n(t,s), m(t,s), l(t,s), j(t,s)\}$, $|s| \leq t < T$, satisfy the initial conditions
\begin{align} \label{GLMinitial}
& m(t,-t) = j(t,-t) = 0,\qquad  n(t,t) = g_0(t), \qquad l(t,t) = \frac{1}{2}|g_0(t)|^2g_0(t) + \frac{i}{2}g_1(t), 
\end{align}
and the ODEs
\begin{align}\label{GLMODEs}
\begin{pmatrix}
n_{t} (t,s) - n_{s}(t,s) \\
m_{t} (t,s) + m_{s}(t,s) \\
l_{t} (t,s)- l_{s}(t,s) \\
j_{t} (t,s) + j_{s}(t,s) 
\end{pmatrix}
=
\begin{pmatrix}
  2i\Delta_2(t) & \alpha_1(t) & 0 & 2g_0(t)\\
  \alpha_2(t) & 0 & \bar{\alpha}_1(t) & 0 \\
  \alpha_3(t) & \alpha_4(t) & 2i\Delta_2(t) & \alpha_5(t) \\
  \alpha_6(t) & \bar{\alpha}_3(t) & \bar{\alpha}_4(t) & 0 
\end{pmatrix}
\begin{pmatrix}
n(t,s) \\
m(t,s) \\
l(t,s) \\
j(t,s)  
\end{pmatrix},
\end{align}
with the functions $\{\alpha_j(t)\}_1^6$ given by
\begin{align*}
& \alpha_1 = \frac{1}{2}\left(|g_0|^2g_0 + 2ig_1\right),
	\\
& \alpha_2 = \frac{1}{4}\left(|g_0|^4\bar{g}_0 + i\bar{g}_0^2g_1 - 2i\bar{g}_{0t}\right),
	\\
& \alpha_3 = \frac{1}{4}\left(i|g_0|^6 + g_0 \bar{g}_{0t} - |g_0|^2 \bar{g}_0g_1 + |g_0|^2g_0\bar{g}_1 + i|g_1|^2\right),
	\\
& \alpha_4 = \frac{1}{8} \left(3 |g_0|^4 g_0 - 2 i g_0^2 \bar{g}_1 + 2 i |g_0|^2 g_1 + 4 i g_{0t}\right),
	\\
& \alpha_5 = |g_0|^2g_0 + ig_1,
	\\
& \alpha_{6} = \frac{1}{32} \Bigl(7|g_0|^6\bar{g}_0 - 4i\bar{g}_0^2 g_{0t} - 12 i |g_0|^2 \bar{g}_{0t} 
	\\
& \hspace{1.6cm}+ 6i|g_0|^2\bar{g}_0^2g_1 - 12 i |g_0|^4 \bar{g}_1 + 8 \bar{g}_0 |g_1|^2 - 4g_0\bar{g}_1^2 - 8\bar{g}_{1t}\Bigr).
\end{align*}
\end{proposition}
\proofbegin
The matrix-valued function $\phi(t,k)$ defined by
\begin{equation}\label{phidef}
  \phi(t,k) = \mu_2(0,t,k) e^{- 2ik^4t\sigma_3}
\end{equation}
satisfies 
\begin{equation}\label{phieq}  
 \phi_t + 2ik^4 \sigma_3 \phi = V_2\phi, \qquad  \phi(0,k) = I.
\end{equation}
We seek a GLM representation for $\phi$ of the form
\begin{equation}\label{phiGLM}
\phi(t,k) = e^{-2ik^4t\sigma_3} + \int_{-t}^t \left[J(t,s) + k L(t,s) + k^{2}M(t,s) + k^{3}N(t,s)\right]e^{-2ik^4s\sigma_3} ds,
\end{equation}
where $L$, $M$, $N$, $J$ are $2 \times 2$-matrix valued functions of $(t,s)$, $|s| \leq t < T$. 
Substitution of (\ref{phiGLM}) into (\ref{phieq}) gives
\begin{align} \label{longJLMN}
& \left[\check{J} + k\check{L} + k^{2}\check{M} + k^{3}\check{N} \right]e^-
+ \left[\hat{J} + k\hat{L} + k^{2}\hat{M} + k^{3}\hat{N} \right]e
	\\ \nonumber
& + \int_{-t}^t \left[J_t + kL_t + k^{2}M_t + k^{3} N_t \right]e^- ds
	+ 2ik^4\sigma_3\int_{-t}^t \left[J + k L + k^{2}M + k^{3}N \right]e^-ds
	\\ \nonumber
& \qquad \qquad \qquad = V_2 \left( e^- + \int_{-t}^t [J + k L + k^{2}M + k^{3}N] e^- ds\right),
\end{align}
where we use the following short-hand notations:
$$e^- := e^{-2ik^4t\sigma_3}, \qquad e := e^{2ik^4t\sigma_3},$$
and
$$\check{F} := F(t,t),Ê\qquad \hat{F} := F(t, -t) \qquad \text{whenever} \qquad \text{$F = N$, $M$, $L$, or $J$};$$
all other occurences of the functions $N,M,L,J$ and their derivatives are understood to be evaluated at $(t,s)$; if the exponential $e^-$ is part of an integrand, it is understood to be evaluated at $s$ instead of at $t$.

We substitute the expression (\ref{V1V2def}) for $V_2$ into equation (\ref{longJLMN}) and apply the integration by parts identity
\begin{align*}
-2ik^4\int_{-t}^t F e^- ds = \biggl[&\check{F}e^- - \hat{F} e
 - \int_{-t}^t F_s e^- ds\biggr]\sigma_3
\end{align*}
to all terms in the resulting equation which are of order $O(k^j)$ with $j \geq 4$.
This yields an equation only containing terms of order $O(k^j)$ with $0 \leq j \leq 3$. Equating the coefficients of the terms of order $k^3$ which do not contain integrals, we find
$$(\check{N} - \sigma_3\check{N}\sigma_3)e^- + (\hat{N} + \sigma_3\hat{N}\sigma_3)e = 2\mathcal{Q}e^-.$$
Equating the coefficients of the terms of order $O(k^3)$ which do contain integrals, we find
\begin{align*}
 \int_{-t}^t N_t e^- ds  +\sigma_3  \int_{-t}^t N_s  e^- ds \sigma_3
 = &\;2\mathcal{Q} \int_{-t}^t J e^- ds -i\mathcal{Q}^2\sigma_3 \int_{-t}^t  L e^- ds
	\\
& + \left(\mathcal{Q}^3 - i\mathcal{Q}_1\sigma_3\right)\int_{-t}^t M e^- ds -i\Delta_2 \sigma_3 \int_{-t}^t N  e^- ds
\end{align*}
We deduce that the terms of order $O(k^3)$ vanish provided that
\begin{align*}
&\check{N} - \sigma_3\check{N}\sigma_3 = 2\mathcal{Q}, \qquad \hat{N} + \sigma_3\hat{N}\sigma_3 = 0,
	\\
& N_t   +\sigma_3 N_s \sigma_3
 = 2\mathcal{Q} J - i\mathcal{Q}^2\sigma_3 L + \left(\mathcal{Q}^3 - i \mathcal{Q}_1\sigma_3\right) M 
 - i\Delta_2 \sigma_3N.
\end{align*} 
Similarly, the terms of $O(k^2)$ vanish provided that
\begin{align*}
& \check{M} - \sigma_3\check{M}\sigma_3 = -i\mathcal{Q}^2\sigma_3 + i\mathcal{Q}\check{N}\sigma_3, \qquad \hat{M} + \sigma_3\hat{M}\sigma_3
= - i\mathcal{Q}\hat{N}\sigma_3,
	\\
& M_t   +\sigma_3 M_s \sigma_3
 = -i\mathcal{Q}^2\sigma_3 J + \left(\mathcal{Q}^3 - i \mathcal{Q}_1\sigma_3\right) L - i\mathcal{Q} N_s\sigma_3 -i\Delta_2 \sigma_3M;
 \end{align*}
the terms of $O(k)$ vanish provided that
\begin{align*}
& \check{L} - \sigma_3\check{L}\sigma_3 =  \mathcal{Q}^3 -i\mathcal{Q}_1\sigma_3 +i\mathcal{Q}\check{M} \sigma_3 + \frac{1}{2}\mathcal{Q}^2\sigma_3\check{N}\sigma_3,
	\\
& \hat{L} + \sigma_3\hat{L}\sigma_3= - i\mathcal{Q}\hat{M} \sigma_3 - \frac{1}{2}\mathcal{Q}^2\sigma_3\hat{N}\sigma_3,
	\\
& L_t   +\sigma_3  L_s \sigma_3 
=  \left(\mathcal{Q}^3 - i \mathcal{Q}_1\sigma_3\right)J - i\mathcal{Q} M_s \sigma_3 - \frac{1}{2}\mathcal{Q}^2\sigma_3 N_s \sigma_3 -i\Delta_2 \sigma_3L;
 \end{align*}
and the terms of $O(1)$ vanish provided that
\begin{align*}
& \check{J} - \sigma_3\check{J}\sigma_3 = i\mathcal{Q}\check{L}\sigma_3 + \frac{1}{2}\mathcal{Q}^2\sigma_3\check{M}\sigma_3 +
\frac{i}{2}(\mathcal{Q}^3-i\mathcal{Q}_1\sigma_3)\check{N}\sigma_3 -i\Delta_2 \sigma_3,
	\\
& \hat{J} + \sigma_3\hat{J}\sigma_3 = - i\mathcal{Q}\hat{L}\sigma_3 
-\frac{1}{2}\mathcal{Q}^2\sigma_3\hat{M}\sigma_3 - \frac{i}{2}(\mathcal{Q}^3-i\mathcal{Q}_1\sigma_3)\hat{N}\sigma_3,
	\\
& J_t   +\sigma_3  J_s  \sigma_3
 =  - i\mathcal{Q} L_s \sigma_3- \frac{1}{2}\mathcal{Q}^2\sigma_3 M_s  \sigma_3  - \frac{i}{2}\left(\mathcal{Q}^3-i\mathcal{Q}_1\sigma_3\right)N_s \sigma_3 -i\Delta_2 \sigma_3J.
 \end{align*}
These equations suggest that we introduce $\tilde{L}$, $\tilde{M}$, $\tilde{J}$ by
\begin{align} \nonumber
& M = \tilde{M} + \frac{i}{2}\mathcal{Q} \sigma_3 N,
	\\ \label{tilderedefinitions}
& L = \tilde{L} + \frac{i}{2}\mathcal{Q}\sigma_3\tilde{M},
	\\ \nonumber
& J = \tilde{J} + \frac{i}{2}\mathcal{Q}\sigma_3\tilde{L} + \left(\frac{i}{8}\mathcal{Q}^3\sigma_3 + \frac{1}{4}\mathcal{Q}_1\right)N.
\end{align}
Straightforward, but tedious, computations show that
\begin{align} \nonumber
& \hat{F} + \sigma_3\hat{F} \sigma_3 = 0 \quad \text{whenever} \quad \text{$F = N$, $\tilde{M}$, $\tilde{L}$, or $\tilde{J}$},
	\\ \nonumber
& N(t,t) - \sigma_3N(t,t)\sigma_3 = 2\mathcal{Q},
	\\ \label{preGLMinitial}
& \tilde{M}(t,t) - \sigma_3\tilde{M}(t,t)\sigma_3 = 0,
	\\ \nonumber
& \tilde{L}(t,t) - \sigma_3\tilde{L}(t,t)\sigma_3 = \mathcal{Q}^3 -i\mathcal{Q}_1\sigma_3,
	\\ \nonumber
& \tilde{J}(t,t) - \sigma_3\tilde{J}(t,t)\sigma_3
= 0,
\end{align}
and
\begin{align} \nonumber
 N_t   +\sigma_3 N_s \sigma_3
 = & \left(\frac{1}{2}\mathcal{Q}^3 - i\mathcal{Q}_1\sigma_3\right)\tilde{M}
+ 2\mathcal{Q} \tilde{J},
 	\\ \nonumber
 \tilde{M}_t + \sigma_3\tilde{M}_s\sigma_3
= & \; \biggl(-\frac{1}{8}\mathcal{Q}^5 - \frac{1}{2}\Delta_2\mathcal{Q} - \frac{i}{2}\mathcal{Q}_t\sigma_3
+ \frac{i}{4}\mathcal{Q}^2\mathcal{Q}_1\sigma_3 \biggr)N
+ \biggl(\frac{1}{2}\mathcal{Q}^3 - i\mathcal{Q}_1\sigma_3\biggr)\tilde{L},
	\\  \nonumber
 \tilde{L}_t + \sigma_3\tilde{L}_s\sigma_3
=  &\;\biggl(\frac{i}{4}\mathcal{Q}^6\sigma_3 - \frac{i}{2}\Delta_2 \mathcal{Q}^2 \sigma_3 + \frac{1}{4}\mathcal{Q}\mathcal{Q}_t
-\frac{1}{8}\mathcal{Q}^2\mathcal{Q}_1\mathcal{Q} + \frac{1}{4}\mathcal{Q}^3\mathcal{Q}_1  - \frac{1}{8}\mathcal{Q}_1\mathcal{Q}^3
	\\ \label{preGLMODEs}
& + \frac{i}{4}\mathcal{Q}_1^2\sigma_3\biggr)N
 + \biggl(- \frac{i}{2}\mathcal{Q}_t\sigma_3 - \frac{1}{2}\Delta_2\mathcal{Q}\biggr)\tilde{M} + \left(\mathcal{Q}^3 - i\mathcal{Q}_1\sigma_3\right)\tilde{J},
	\\ \nonumber
 \tilde{J}_t + \sigma_3\tilde{J}_s\sigma_3
= & -\frac{i}{2}\mathcal{Q}\sigma_3\left(\tilde{L}_t + \sigma_3\tilde{L}_s\sigma_3\right)
+ \left(-\frac{i}{8}\mathcal{Q}^3\sigma_3 - \frac{1}{4}\mathcal{Q}_1\right)\left(N_t + \sigma_3N_s\sigma_3\right) 
	\\ \nonumber
 & - \frac{i}{2}\mathcal{Q}_t\sigma_3\tilde{L} - \frac{i}{8}(\mathcal{Q}^3)_t\sigma_3N - \frac{1}{4}\mathcal{Q}_{1t} N -i\Delta_2 \sigma_3J.
\end{align}
The symmetries
$$\phi(t,k) = \sigma_3\phi(t,-k)\sigma_3, \qquad \phi(t, k) = \sigma_1\overline{\phi(t,\bar{k})}\sigma_1,$$
suggest that we seek $L$, $\tilde{M}$, $\tilde{N}$, $\tilde{J}$ such that $\tilde{J}$ and $\tilde{M}$ (resp. $L$ and $\tilde{N}$) are diagonal (resp. off-diagonal) matrices, and such that
$$F = \sigma_1\bar{F} \sigma_1 \quad \text{whenever} \quad \text{$F = N$, $\tilde{M}$, $\tilde{L}$, or $\tilde{J}$}.$$
In fact, introducing complex-valued functions $\{n(t,s), m(t,s), l(t,s), j(t,s)\}$ by
\begin{align} \nonumber
&N = \begin{pmatrix} 0 & ne^{-2i\int_{(0,0)}^{(0,t)} \Delta } \\ \bar{n}e^{2i\int_{(0,0)}^{(0,t)} \Delta } & 0 \end{pmatrix}, \qquad
\tilde{M} = \begin{pmatrix} \bar{m} & 0 \\ 0 & m \end{pmatrix},
	\\ \label{nmlkdef}
&\tilde{L} = \begin{pmatrix} 0 & le^{-2i\int_{(0,0)}^{(0,t)} \Delta} \\ \bar{l}e^{2i\int_{(0,0)}^{(0,t)} \Delta} & 0 \end{pmatrix}, \qquad
\tilde{J} = \begin{pmatrix} \bar{j} & 0 \\ 0 & j \end{pmatrix},
\end{align}
a long computation shows that the two sets of equations (\ref{preGLMinitial}) and (\ref{preGLMODEs}) are satisfied provided that $\{n, m, l, j\}$ satisfy the initial conditions (\ref{GLMinitial}) and the ODEs (\ref{GLMODEs}), respectively.

Equations (\ref{phiGLM}), (\ref{tilderedefinitions}), and (\ref{nmlkdef}) imply that
\begin{align*}
& \begin{pmatrix} e^{2i\int_{(0,0)}^{(0,t)} \Delta}\phi_{12}(t,k) \\
 \phi_{22}(t,k) \end{pmatrix}
  = \begin{pmatrix} 0 \\ e^{2ik^4t} \end{pmatrix}
+  \int_{-t}^t \begin{pmatrix} I_1(t,s)k + I_3(t,s) k^3 \\ I_0(t,s) + I_2(t,s) k^2 \end{pmatrix} e^{2ik^4s} ds,
\end{align*}
where $\{I_j\}_0^3$ are given by (\ref{I0123def}). The relations
$$\phi_{12} = \Phi_1e^{-2i\int_0^t\Delta_2} e^{2ik^4t}, \qquad
\phi_{22} = \Phi_2e^{2ik^4t}$$
then imply that $\{\Phi_j\}_1^2$ are given by (\ref{GLMrep}).
\proofend

\section{The solution of the global relation: physical domain}\nequation\label{physicalsolutionsec}
In this section, we consider the solution of the global relation in the physical domain. Theorems \ref{th2} and \ref{th3} below lead to expressions for the Dirichlet-to-Neumann map in terms of the solution of a system of nonlinear integral equations. Theorem \ref{th2} applies to arbitrary initial data. Theorem \ref{th3} provides a more convenient expression for the Dirichlet-to-Neumann map in the special, but important, case of vanishing initial data.

\subsection{The general case}\label{generalsubsec}
For a function $F(t,s)$, $-t < s < t$, we let $\mathcal{A}$ denote the Abel transform in the second variable defined by
$$(\mathcal{A}F)(t,s) = \int_{-t}^s F(t,s') \frac{ds'}{\sqrt{s - s'}}.$$
The inverse $\mathcal{A}^{-1}$ of this transform is given by
$$(\mathcal{A}^{-1}F)(t,s) = \frac{1}{\pi} \frac{d}{ds}\int_{-t}^s \frac{F(t,s')ds'}{\sqrt{s - s'}} = \frac{1}{\pi}\left[\int_{-t}^s \frac{\frac{\partial F}{\partial s'}(t,s')ds'}{\sqrt{s-s'}} + \frac{F(t,-t)}{\sqrt{t+s}}\right].$$

\begin{theorem}\label{th2}
Let $q_0(x)$, $g_0(t)$, $a(k)$, and $b(k)$ be as in theorem \ref{th1} and let $\{k_j, -k_j\}_1^{N} \subset D_1 \cup D_2$ denote the possible simple zeros of $a(k)$. 
Define functions $f_0(t)$ and $f_2(t)$, $t > 0$, by
\begin{equation}\label{fjdef}
f_j(t) = \int_{\gamma'} k^j \frac{b(k)}{a(k)}e^{-2ik^4t} dk,  \qquad j = 0, 2,\quad t > 0,
\end{equation}
where $\gamma'$ is the contour defined in section \ref{spectralsolutionsec}.
Let $q(x,t)$, $x > 0$, $0 < t < T$, be a solution of the DNLS equation (\ref{dnls}) which satisfies
$$q(x, 0) = q_0(x), \quad x > 0; \quad q(0,t) = g_0(t), \quad 0 < t < T.$$
Let $\{n,m,l,j\}$ be the functions in the GLM representation (\ref{GLMrep}).

Then the Neumann boundary data $g_1(t) := q_x(0, t)$ are given by
\begin{align} \nonumber
  g_1(t)& = i|g_0(t)|^2g_0(t) - 4c_1(\mathcal{A}^{-1}n)(t,t) + g_0(t)m(t,t) +   \frac{4e^{2i\int_{(0,0)}^{(0,t)} \Delta}}{\pi}\biggl\{2if_2(2t) 
  	\\ \nonumber
& + 2i \int_{-t}^t \biggl[ \bar{j}(t,s) - \frac{i}{2}g_0(t)\bar{l}(t,s) 
 + \left(\frac{i}{8}|g_0(t)|^2g_0(t) + \frac{1}{2i}l(t,t)\right)\bar{n}(t,s)\biggr] f_2(t + s)ds
  	\\ \label{g1th2}
&- \left(\bar{m}(t,t) - \frac{i}{2}|g_0(t)|^2\right)f_0(2t) 
+ \frac{\pi i}{16}g_0(t)\bar{n}(t,-t) q_0(0)
	\\ \nonumber
& + \int_{-t}^t \left(\bar{m}_s(t,s) - \frac{i}{2}g_0(t)\bar{n}_s(t,s)\right) f_0(t +s)ds\biggr]\biggr\},
  \end{align}
where $c_1 \in \C$Ê is a constant defined by
\begin{equation}\label{c1def}  
  c_1 = \int_{\gamma} l e^{-2il^4}dl = \frac{\sqrt{\pi}e^{-i\pi/4}}{2\sqrt{2}}.
\end{equation}
\end{theorem}
\proofbegin
Substituting the GLM representation (\ref{GLMrep}) into the global relation (\ref{GRc}), we find the following equation which is valid for $k \in \bar{D}_1 \cup \bar{D}_2$:
\begin{align*}
&  \int_{-t}^t \left( I_1(t,s')k + I_3(t,s') k^3\right) e^{2ik^4(s' -t)} ds'
	\\
&  + e^{2i\int_{(0,0)}^{(0,t)} \Delta} \frac{b(k)}{a(k)}\left[1 + \int_{-t}^t \left( \bar{I}_0(t,s') + \bar{I}_2(t,s') k^2\right) e^{-2ik^4(s' -t)} ds' \right]e^{-4ik^2t} = c(t, k).
\end{align*}
We multiply this equation by $k^2e^{2ik^4(t-s)}$, $-t < s < t$, and integrate the resulting equation with respect to $dk$ along $\gamma'$. This yields
\begin{align} \label{GRGLMidentity}
& \int_{\gamma'} k^2 \int_{-t}^t \left( I_1(t,s')k + I_3(t,s') k^3\right) e^{2ik^4(s' - s)} ds' dk + e^{2i\int_{(0,0)}^{(0,t)} \Delta}
	\\ \nonumber
&\qquad \times \int_{\gamma'} k^2\frac{b(k)}{a(k)}\left[e^{-2ik^2(t+s)} + \int_{-t}^t \left( \bar{I}_0(t,s') + \bar{I}_2(t,s') k^2\right) e^{-2ik^4(s' + s)} ds' \right]dk = 0,
\end{align}
where the right-hand side is zero because of Jordan's lemma. Analyticity implies that the contour $\gamma'$ can be replaced by $\gamma$ in the first integral in (\ref{GRGLMidentity}).
Using the Fourier inversion identity\footnote{This identity can be less rigorously expressed as 
$\int_{\gamma} k^3 e^{2ik^4(s' -s)}dk = \frac{\pi}{4}\delta(s'-s).$}
\begin{equation}\label{intidentity1}
\int_{\gamma} k^3 \left[\int_{-t}^t K(t, s') e^{2ik^4(s' - s)}ds'\right] dk = \begin{cases} \frac{\pi}{4}K(t, s), \qquad -t < s < t, \\
\frac{\pi}{8}K(t, t), \qquad 0 < s = t, \end{cases}
\end{equation}
we see that the term in (\ref{GRGLMidentity}) involving $I_1$ equals
\begin{equation}\label{I1term}
\int_{\gamma'} k^3 \int_{-t}^t I_1(t,s')e^{2ik^4(s' - s)} ds' dk
= \frac{\pi}{4}I_1(t,s).
\end{equation}
In order to simplify the term in (\ref{GRGLMidentity}) involving $I_3$, we integrate by parts as follows:
\begin{align*}
&\int_{\gamma} k^5 \int_{-t}^t I_3(t,s') e^{2ik^4(s' - s)} ds' dk
	\\
&= \frac{1}{2i}\int_{\gamma} k \left[I_3(t,t)e^{2ik^4(t-s)} - I_3(t, -t)e^{-2ik^4(t+s)} - \int_{-t}^t I_{3s'}(t,s') e^{2ik^4(s' - s)} ds'\right] dk.\end{align*}
We can now interchange the order of integration in the integral on the right-hand side involving $I_{3s'}$. Using the identity
\begin{equation}\label{intidentity2}
\int_{\gamma} k e^{-2ik^4 t}dk = \begin{cases} 0, \qquad t < 0, \\
\frac{c_1}{\sqrt{t}}, \qquad t > 0,\end{cases}
\end{equation}
where $c_1$ is given by (\ref{c1def}),
we find that the term in (\ref{GRGLMidentity}) involving $I_3$ equals
\begin{equation}\label{I3term}
- \frac{c_1}{2i} \left[\frac{I_3(t, -t)}{\sqrt{t+s}} + \int_{-t}^s I_{3s'}(t,s') \frac{ds'}{\sqrt{s - s'}} \right]
= - \frac{c_1\pi}{2i}(\mathcal{A}^{-1}I_3)(t,s).
\end{equation}

In view of the definition (\ref{fjdef}) of $f_0$ and $f_2$, the term in (\ref{GRGLMidentity}) involving $\bar{I}_0$ equals
\begin{equation}\label{I0term}
  e^{2i\int_{(0,0)}^{(0,t)} \Delta}  \int_{-t}^t \bar{I}_0(t,s') f_2(s+s') ds'.
\end{equation}  
In order to simplify the term in (\ref{GRGLMidentity}) involving $\bar{I}_2$, we integrate by parts as follows:
\begin{align*}
 \int_{\gamma'} k^4\frac{b(k)}{a(k)} \int_{-t}^t \bar{I}_2(t,s') e^{-2ik^4(s' + s)} &ds' dk 
	\\
=  -\frac{1}{2i}\int_{\gamma'} \frac{b(k)}{a(k)}  \biggl[ &\bar{I}_2(t,t) e^{-2ik^4(t + s)} - \bar{I}_2(t, -t) e^{-2ik^4(-t + s)} 
	\\
& - \int_{-t}^t \bar{I}_{2s'}(t,s') e^{-2ik^4(s' + s)} ds' \biggr]  dk.
\end{align*}
Expressing the integrals with respect to $dk$ on the right-hand side in terms of $f_0$, we infer that the term in (\ref{GRGLMidentity}) involving $\bar{I}_2$ equals
\begin{equation}\label{I2term}
  -\frac{e^{2i\int_{(0,0)}^{(0,t)} \Delta} }{2i}\left[\bar{I}_2(t,t)f_0(t+s) - \bar{I}_2(t,-t)f_0(-t+s) - \int_{-t}^t \bar{I}_{2s'}(t,s')f_0(s'+s) ds'\right].
\end{equation}  

Using equations (\ref{I1term}) and (\ref{I3term})-(\ref{I2term}), we can write (\ref{GRGLMidentity}) as
\begin{align*} \label{}
 &\frac{\pi}{4}I_1(t,s) - \frac{c_1\pi}{2i}(\mathcal{A}^{-1}I_3)(t,s)
+ e^{2i\int_{(0,0)}^{(0,t)} \Delta}\biggl\{ f_2(t+s) +  \int_{-t}^t \bar{I}_0(t,s') f_2(s+s') ds'
	\\
&-\frac{1}{2i}\left[\bar{I}_2(t,t)f_0(t+s) - \bar{I}_2(t,-t)f_0(-t+s) - \int_{-t}^t \bar{I}_{2s'}(t,s')f_0(s'+s) ds'\right]\biggr\}
= 0,
\end{align*}
Recalling the definitions (\ref{I0123def}) of $\{I_j\}_0^3$ and the initial conditions (\ref{GLMinitial}), evaluation of this equation as Ê$s \to t$ yields
\begin{align} \nonumber
& \frac{\pi}{4}\biggl(\frac{1}{2}|g_0(t)|^2g_0(t) + \frac{i}{2}g_1(t) - \frac{i}{2}g_0(t)m(t,t)\biggr) 
 - \frac{c_1\pi}{2i}(\mathcal{A}^{-1}n)(t,t) +  e^{2i\int_{(0,0)}^{(0,t)} \Delta} \biggl\{f_2(2t) 
	\\ \nonumber
& + \int_{-t}^t 
\biggl[ \bar{j}(t,s) - \frac{i}{2}g_0(t)\bar{l}(t,s) 
+ \left(\frac{i}{8}|g_0(t)|^2g_0(t) + \frac{1}{2i}l(t,t)\right)\bar{n}(t,s)\biggr] f_2(t+s) ds
	\\  \label{preg1th2}
&-\frac{1}{2i}\biggl[\left(\bar{m}(t,t) - \frac{i}{2}|g_0(t)|^2\right)f_0(2t) 
+ \frac{i}{2}g_0(t)\bar{n}(t,-t) f_0(0) 
	\\ \nonumber
&- \int_{-t}^t \left(\bar{m}_s(t,s) - \frac{i}{2}g_0(t)\bar{n}_s(t,s)\right)f_0(t+s) ds\biggr]\biggr\}
= 0,
\end{align}
Since
$$\frac{b(k)}{a(k)} = - \frac{iq_0(0)}{2k} + O\biggl(\frac{1}{k^3}\biggr), \qquad k \to \infty, \quad k \in D_1 \cup D_2,$$
we find
\begin{equation}\label{f00expression}
f_0(0) =  \int_{\gamma'} \frac{b(k)}{a(k)} dk
= - \int_{\gamma'} \frac{iq_0(0)}{2k} dk
= - \frac{\pi q_0(0)}{8}.
\end{equation}
Solving equation (\ref{preg1th2}) for $g_1(t)$ and using (\ref{f00expression}), we find (\ref{g1th2}).
\proofend

Substitution of the expression (\ref{g1th2}) for $g_1(t)$ into (\ref{GLMODEs}) yields a nonlinear system of integro-differential equations involving $\{n,m,l,j\}$ and $\Delta_2$. 
Employing the identities
\begin{align}\label{Fidentities}
& (\partial_t - \partial_s)F(t,s) = f(t,s) \implies F(t,s) = F \Bigl(\frac{t+s}{2}, \frac{t+s}{2} \Bigr) + \int_{\frac{t+s}{2}}^t f\left(\tau, t+s-\tau\right)d\tau,
	\\ \nonumber
& (\partial_t + \partial_s)F(t,s) = f(t,s) \implies F(t,s) = F\Bigl(\frac{t-s}{2}, -\frac{t-s}{2} \Bigr) + \int_{\frac{t-s}{2}}^t f(\tau, \tau - t + s)d\tau,
\end{align}
and recalling the initial conditions in (\ref{GLMinitial}), we can convert this into a system of nonlinear integral equations. Supplementing this system with the equation for $\Delta_2$ obtained by substituting  the expression (\ref{g1th2}) for $g_1(t)$ into the equation
$$\Delta_2(0,t) = \frac{3}{4}|g_0|^4 - \frac{i}{2}(\bar{g}_1g_0 - \bar{g}_0g_1),$$ 
we find a closed system of nonlinear integral equations for the complex-valued functions $\{n(t,s),m(t,s),l(t,s),j(t,s)\}$, $|s| \leq t < T$ and the real-valued function $\Delta_2(0,t)$, $0 < t < T$. This system is formulated explicitly in terms of the initial data $q_0(x)$ and the Dirichlet boundary data $g_0(t)$. Once the solution of this system has been found,\footnote{The problem of rigorously showing existence and uniqueness for this system remains open.} the Neumann data $g_1(t)$ can be determined from (\ref{g1th2}). This provides an effective way of computing the Dirichlet-to-Neumann map.

\begin{remark}\upshape
The expression (\ref{g1th2}) for $g_1$ in terms of $\{n,m,l,j\}$ can also be derived directly from the expression (\ref{g1expression}) for $g_1$ given in theorem \ref{th1}. Indeed, (\ref{g1expression}) states that $g_1$ is given by
\begin{equation}\label{g1remark}
g_1= \frac{i}{2}|g_0|^2g_0 + 4 c^{(3)} + 2ig_0 \Phi_2^{(2)},
\end{equation}
where $c^{(3)}$ is given by (\ref{c3expression}) and 
\begin{equation}\label{Phi2inrem}  
  \Phi_2^{(2)} = \frac{2}{\pi i}\int_{\gamma} k [\Phi_2(t,k) - \Phi_2(t, ik)] dk.
\end{equation}
Substituting the GLM representation (\ref{GLMrep}) for $\Phi_2$ into (\ref{Phi2inrem}) and using the identity (\ref{intidentity1}), we conclude that
$$2ig_0\Phi_2^{(2)} = \frac{8g_0}{\pi} \int_{\gamma} k^3\int_{-t}^t I_2(t,s)e^{2ik^4(s - t)}ds dk
= g_0\left(m(t,t) + \frac{i}{2}\bar{g}_0n(t,t)\right).$$
On the other hand, we can write $c^{(3)}$ as the sum of two terms, $c^{(3)} = c_1^{(3)} +c_2^{(3)}$, where 
\begin{equation}\label{c13def}  
  c_1^{(3)}= \frac{i}{\pi}\int_{\gamma} k^2\left(\Phi_1(t,k) + i \Phi_1(t, ik) + \frac{ig_0}{k}\right) dk,
\end{equation}
and $c_2^{(3)}$ involves the quotient $b(k)/a(k)$. Substituting the GLM representation for $\Phi_1$ into (\ref{c13def}) and integrating the term in the resulting expression involving $k^5$ by parts, we find
$$c_1^{(3)} = \frac{1}{\pi}\int_{\gamma} \left[kn(t,t) - kn(t,-t)e^{-4ik^4t} - \int_{-t}^t kn_{s}(t,s)e^{2ik^4(s-t)}ds -kg_0\right]dk.$$
Using the identity (\ref{intidentity2}) and the initial condition $n(t,t) = g_0(t)$, this gives
\begin{equation}\label{c13final}  
  c_1^{(3)} = -c_1 (\mathcal{A}^{-1}n)(t,t).
\end{equation}
Finally, simplifying $c_2^{(3)}$ as in the proof of theorem \ref{th2}, we deduce that $c_2^{(3)}$ is given by the terms on the right-hand side of equation (\ref{g1th2}) which involve the functions $f_0$ and $f_2$.
In summary, after collecting the various contributions, we conclude that (\ref{g1th2}) can be derived from (\ref{g1expression}).
\end{remark}

\subsection{The case of vanishing initial data}
With respect to the integral equations derived in the spectral domain (see theorem \ref{th1}), the system of integral equations derived in section \ref{generalsubsec} has the numerical advantage of being defined on the {\it bounded} domain $|s| \leq t < T$. Nevertheless, the system of section \ref{generalsubsec} involves five highly nonlinear and rather intricate equations. The next theorem shows that this system can be simplified considerably in the case of vanishing initial data $q_0(x) = 0$. 

\begin{theorem}\label{th3}
Suppose that $q_0(x) = 0$, $x \geq 0$. Let $g_0(t)$ and $q(x,t)$ be as in theorem \ref{th2}. Then the Neumann boundary data $g_1(t) := q_x(0,t)$ are given by
\begin{equation}\label{g1th3}
  g_1(t) = i|g_0(t)|^2g_0(t) - 4c_1(\mathcal{A}^{-1}n)(t,t) + g_0(t)m(t,t),
\end{equation}
where $c_1$ is given by (\ref{c1def}) and the functions $\{n(t,s), m(t,s)\}$, $|s| \leq t < T$, satisfy the following system of two nonlinear integral equations:
\begin{subequations}\label{nmsystem}
\begin{align}\label{nmsystema}
& n(t,s) = g_0\biggl(\frac{t+s}{2}\biggr)
+ \int_{\frac{t +s}{2}}^t \left[\beta_1(\tau)n + \beta_2(\tau)m + \beta_3(\tau)(\mathcal{A}^{-1}m)\right] d\tau,
	\\ \label{nmsystemb}
& m(t,s) =  \int_{\frac{t - s}{2}}^t \left[\beta_4 (\tau) m + \beta_5 (\tau) n + \beta_6 (\tau) (\mathcal{A}^{-1}n)\right] d\tau,
\end{align}
\end{subequations}
with
\begin{align}\nonumber
& \beta_1 = \frac{5i}{4}|g_0|^4 + \frac{1}{2}g_0\bar{g}_1 - \bar{g}_0 g_1,
	\\ \nonumber
& \beta_2 = |g_0|^2g_0 + i g_1,
	\\ \nonumber
& \beta_3 = -\frac{\pi g_0}{2c_1},
	\\ \label{bjdef}
& \beta_4 = \frac{1}{4}\left(i|g_0|^4 + 2g_0\bar{g}_1\right),
	\\ \nonumber
& \beta_5 = \frac{1}{4}\left(|g_0|^4\bar{g}_0 + i \bar{g}_0^2 g_1 - 2i\bar{g}_{0t} \right),
	\\ \nonumber
& \beta_6 = -\frac{\pi}{8 c_1}\left(|g_0|^2\bar{g}_0 - 2i\bar{g}_1\right).
\end{align}
Unless otherwise specified, the functions in (\ref{nmsystema}) are evaluated at $(\tau, t+s-\tau)$ and the functions in (\ref{nmsystemb}) are evaluated at $(\tau, \tau - t + s)$.
\end{theorem}
\proofbegin
The expression (\ref{g1th3}) for $g_1(t)$ is obtained by setting $f_0(k) = f_2(k) = q_0(0) = 0$ in (\ref{g1th2}).
We will use the global relation to eliminate the functions $l$ and $j$ from the equations (\ref{GLMODEs}) for $\{n,m,l,j\}$; this will lead to (\ref{nmsystem}). 

Substituting the GLM representations (\ref{GLMrep}) into the global relation (\ref{GRcd}) and using that $b(k) = 0$, we find
\begin{subequations}
\begin{align}\label{GLMinglobala}
 & \int_{-t}^t (kI_1 + k^3I_3) e^{2ik^4(s' -t)} ds' = c(k,t),
 	\\ \label{GLMinglobalb}
& \int_{-t}^t (I_0 + I_2k^2) e^{2ik^4(s' -t)} ds' = d(k,t) - 1,
\end{align}
\end{subequations}
where $c$ and $d - 1$ are analytic functions of $k \in D_1 \cup D_2$ and of order $O(1/k)$ as $k \to \infty$.

We first analyze (\ref{GLMinglobala}).
We multiply (\ref{GLMinglobala}) by $e^{2ik^4(t -s)}$ with $-t < s < t$ and integrate the resulting equation with respect to $dk$ along $\gamma$. 
This yields
\begin{equation}\label{I1I3first}
  \int_{\gamma} \int_{-t}^t (kI_1 + k^3I_3)(t, s') e^{2ik^4(s' - s)} ds' dk = 0,
\end{equation}  
where the contribution of the right-hand side vanishes by Jordan's lemma.
Using the identities (\ref{intidentity1}) and (\ref{intidentity2}), equation (\ref{I1I3first}) becomes
\begin{equation}\label{I1I3second}
  c_1(\mathcal{A}I_1)(t,s) + \frac{\pi}{4} I_3(t, s) = 0, \qquad -t < s < t.
\end{equation}
Applying the inverse Abel transform to (\ref{I1I3second}) and using the expressions  for $I_1$ and $I_3$ given in (\ref{I0123def}), we find\footnote{In view of (\ref{GLMinitial}), evaluation of (\ref{I1I3third}) as $s \to t$ yields an alternative derivation of (\ref{g1th3}).}
\begin{align} \label{I1I3third}
  c_1 \left[l(t,s) - \frac{i}{2}g_0(t)m(t,s)\right] + \frac{\pi}{4} (\mathcal{A}^{-1}n)(t, s) = 0, \qquad -t < s < t.
\end{align}
Similarly, we multiply (\ref{GLMinglobalb}) by $ke^{2ik^4(t -s)}$ with $-t < s < t$ and integrate the resulting equation with respect to $dk$ along $\gamma$. Steps analogous to those that led to (\ref{I1I3third}) now yield
\begin{align}\label{I1I3thirdanalog}
& c_1 \left[j(t,s) + \frac{i}{2}\bar{g}_0(t)l(t,s) + \left(\frac{i}{8}|g_0(t)|^2\bar{g}_0(t) + \frac{1}{4}\bar{g}_1(t)\right)n(t,s) \right] 
	\\\nonumber
&+ \frac{\pi}{4} (\mathcal{A}^{-1}m)(t,s) + \frac{\pi i}{8}\bar{g}_0(t)(\mathcal{A}^{-1}n)(t, s) = 0, \qquad -t < s < t.  
\end{align}
We solve the equations (\ref{I1I3third}) and (\ref{I1I3thirdanalog}) for $j$ and $l$ and plug the result into the first two equations of (\ref{GLMODEs}). This yields
$$\begin{pmatrix}
n_{t} - n_{s} \\
m_{t} + m_{s} 
\end{pmatrix}
=
\begin{pmatrix} \beta_1 & \beta_2 + \beta_3\mathcal{A}^{-1}  \\
\beta_5 + \beta_6\mathcal{A}^{-1}  & \beta_4 
\end{pmatrix}
\begin{pmatrix} n \\ m  \end{pmatrix},
$$
where $\{\beta_j\}_1^6$ are given by (\ref{bjdef}).
Using the identities (\ref{Fidentities}) and the initial conditions (\ref{GLMinitial}), we find (\ref{nmsystem}).
\proofend

Substitution of the expression (\ref{g1th3}) for $g_1(t)$ into (\ref{nmsystem}) yields a system of nonlinear integral equations for $\{n(t,s),m(t,s)\}$, $|s| \leq t < T$, formulated entirely in terms of the known boundary values $g_0(t)$. This system only involves two equations for two unknown functions. This is in contrast to the system derived in section \ref{generalsubsec} in the case of general initial data, which involves five equations.

\bigskip
\noindent
{\bf Acknowledgement} {\it The author is grateful to A. S. Fokas for helpful discussions.}

\bibliographystyle{plain}
\bibliography{is}

\end{document}